\documentclass[%
 twocolumn, 
 amsmath,amssymb,
 aps,
prb,
]{revtex4-2}
\usepackage{graphicx}% Include figure files
\usepackage{dcolumn}% Align table columns on decimal point
\usepackage{bm}% bold math
\usepackage{threeparttable} % for tablenotes environment
\usepackage{color} 
\usepackage{comment}
\usepackage{threeparttable}
\usepackage[capitalize]{cleveref}
\usepackage{tablefootnote}
\usepackage[T1]{fontenc}
\usepackage[utf8]{inputenc} % if using pdflatex
\usepackage{lmodern}
\usepackage[normalem]{ulem}
\newcommand\blue[1]{{\color{blue}#1}}
\newcommand\red[1]{{\color{red}#1}}

\definecolor{green}{rgb}{0.0, 0.5, 0.0}

\begin{document}
\preprint{APS/123-QED}
 
%\begin{frontmatter}
%\title{Electric field control the magnetism in bilayer VI_{3}}%
\title{%Dirac half-metal 
Quantum anomalous Hall effect in monolayer transition-metal trihalides} %Force line breaks with \\



\author{Thi Phuong Thao Nguyen}
\affiliation{%
 SANKEN, The University of Osaka, 8-1 Mihogaoka, Ibaraki, Osaka, 567-0047, Japan
}%
\author{Kunihiko Yamauchi}
\affiliation{%
 Center for Spintronics Research Network (CSRN), The University of Osaka, 1-3 Machikaneyamacho, Toyonaka, Osaka 560-8531, Japan
}%

\date{\today}% It is always \today, today,

\begin{abstract}
We present systematic first-principles results for the electronic and magnetic properties of two-dimensional transition-metal trihalide monolayers, $MX_3$ ($M$ = V, Cr, Mn, Fe, Ni, Pd; $X$ = F, Cl, Br, I), focusing on their potential to host the quantum anomalous Hall effect. In particular, MnF$_3$ and PdF$_3$ exhibit a spin-polarized Dirac cone at the $K$ point; spin–orbit coupling opens a sizable gap with a nonzero Chern number $C=\pm 1$. Nanoribbon slab calculations reveal gap-crossing chiral edge states, establishing the nontrivial topological character. Beyond these case studies, our systematic screening clarifies general trends across the $MX_3$ family and provides insight into how electronic configuration and spin–orbit coupling cooperate to produce magnetic and topological phases in two-dimensional magnets.
\end{abstract}

\maketitle

%\end{frontmatter}
%\linenumbers
\begin{comment}

\blue{-Arrange figures so that we can see text and figures together in the same page.} 

\red{
- TN updated all the figures and tried to arrange them with the text. It seems like we have more figures than text.}

\blue{- Try to reduce number of LaTeX compile errors.

- I don't like your use of "correlation" term. So I deleted them from text. }
    
\end{comment}

\section{Introduction}

\begin{figure}[!b]
%\begin{center}
%\vspace{-0.3cm}
%\resizebox{120mm}{!}
{
 \includegraphics[width=90mm, angle=0]{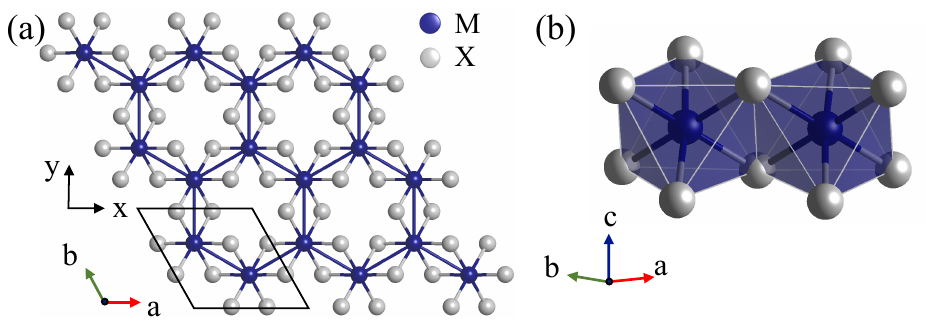}
%\vspace {-0.4cm}
}
\caption{\label{fig:model}
(a) Top view of $MX_3$ monolayer structure. The black line shows the hexagonal unit cell. Transition-metal atoms form a honeycomb-pattern layered lattice.
(b) Schematic illustration of the $MX_6$ octahedron, where each $M^{3+}$ ion is coordinated by six $X^-$ ions.
} 
%\vspace {-1.0cm}
%\end{center}
\end{figure}

\begin{comment}

\begin{tablenotes}
    \cite{He2016} J. He (2016)  GGA+$U$ ; \cite{Nguyen2021} T.P.T Nguyen (2021) GGA+$U$; \cite{Tomar2019} S. Tomar (2019) HSE ; \cite{PingLi2019} P. Li (2019) HSE; \cite{QSun2018} Q. Sun (2018) GGA+$U$;  \cite{JHe2017_nicl3} J. He (2017) HSE;  \cite{Wang2018} Y-P. Wang (2018) GGA+$U$; 
 \cite{Bafekry2020} A. Bafekry (2020) GGA+$U$; \cite{ZhengLi2019} Z. Li (2019) HSE;   \cite{Olsen2019} T. Olsen (2019) GGA; \cite{You2019} J-Y. You  (2019) HSE. 
 
    Note: In this work, \textit{Dirac half-metal} (DHM) is defined as a material in which one spin channel exhibits a Dirac-like linear crossing at the Fermi level, while the opposite spin channel has a finite band gap and is insulating. \textit{QAHI} denotes a quantum anomalous Hall insulator, i.e., a gapped system with a finite Chern number.  
\textit{Ins.} refers to a trivial (non-topological) insulator.  
In contrast, \textit{SM} (semimetal) is defined as a system where valence and conduction bands overlap slightly near the Fermi level, with both spin channels contributing to electronic conduction. 
\red{(TN: This table note is lengthy, we can remove the first part and move a note about Ins., SM, DHM.... to the main text. ==> move it to footnote)}
\end{tablenotes}
\end{comment}

The quantum anomalous Hall effect (QAHE), characterized by quantized Hall conductance without an external magnetic field, 
has become a central theme in condensed matter physics owing to its profound connection to band topology 
and its potential in spintronics and quantum information technologies\cite{Chang2013_QAH, Chang2023_RMP}. 
Since its first experimental realization in magnetically doped topological insulators, 
the search for robust QAHE systems has been an active frontier, 
motivating efforts to identify intrinsic magnetic materials that can host this phase at elevated temperatures. 
Two-dimensional (2D) materials, with their reduced dimensionality and the interplay of $d$-electron configuration, magnetism, and spin–orbit coupling (SOC), provide a particularly promising platform for realizing the QAHE\cite{Zhang2024_PRB}.

Among them, monolayer transition-metal trihalides $MX_3$ ($M$: transition-metal ion; $X$: halide ion) have recently attracted considerable interest. Their crystal structure consists of transition-metal atom arranged in a honeycomb pattern, each octahedrally coordinated by six halogen atoms, as illustrated in Fig. \ref{fig:model}. This geometry forms edge-sharing $MX_6$ octahedra, a motif characteristic of many layered magnetic halides such as CrI$_3$, %The combination of electron-electron correlations, and strong spin–orbit coupling (SOC) makes the monolayer transition-metal trihalides natural candidates for topological magnetic phases.
%In addition to their structural appeal, $MX_3$ monolayers have been widely studied as 
which is now recognized as a prototype of 2D magnets. 
For example, ferromagnetism in monolayer CrI$_3$ was experimentally demonstrated\cite{Huang2017}, 
interlayer antiferromagnetic coupling was reported in bilayer CrI$_3$\cite{Song2018}, 
and electric-field control of magnetic order has also been achieved\cite{Jiang2018}. 
These works highlight the versatility of $MX_3$ compounds as platforms for exploring tunable 2D magnetism.

Despite these advances and the growing interest in $MX_3$ monolayers as versatile 2D magnets, 
their intrinsic electronic and magnetic ground states remain under debate. 
Theoretical studies have reported diverse and sometimes conflicting results (see Table \ref{table:overview}) \footnote[1]{ In this work, \textit{Dirac half-metal} (DHM) is defined as a material in which one spin channel exhibits a Dirac-like linear crossing at the Fermi level, while the opposite spin channel has a finite band gap and is insulating. \textit{QAHI} denotes a quantum anomalous Hall insulator, i.e., a gapped system with a finite Chern number.  \textit{Ins.} refers to a trivial (non-topological) insulator.  In contrast, \textit{SM} (semimetal) is defined as a system where valence and conduction bands overlap slightly near the Fermi level, with both spin channels contributing to electronic conduction. }. For example, \citet{He2016} predicted QAHE in VCl$_3$ and VI$_3$ monolayers through SOC-induced band gaps with nonzero Chern numbers. However, \citet{Zhou2016} proposed that VCl$_3$ is half-metallic, while \citet{Zhao2021} identified it as an antiferromagnetic semiconductor. For VI$_3$, metallic behavior has been reported in several works \citep{vi3_ml2_metallic, vi3_ml1_2states}, whereas \citet{Nguyen2021} and others \citep{vi3_ml3_insu_soc, vi3_stack, vi3_cri3_arpes, vi3_ml_insu_dop} demonstrated an insulating ground state. Similarly, FeX$_3$ was found to be antiferromagnetic and insulating in Ref.\cite{Tomar2019}, while Ref.\cite{PingLi2019} suggested a spin-polarized Dirac half-metal. Other studies reported the possibility of QAHE in MnX$_3$, NiCl$_3$, and PdCl$_3$\cite{QSun2018,JHe2017_nicl3,Wang2018,Bafekry2020}.

These conflicting reports indicate that the ground states and topological properties of transition-metal trihalide monolayers remain unsettled. A systematic investigation across different chemical compositions is therefore needed to clarify how their electronic and magnetic behaviors evolve. In this work, we performed comprehensive first-principles calculations for representative $MX_3$ monolayers ($M$ = V, Cr, Mn, Fe, Ni, Pd; $X$ = F, Cl, Br, I) to elucidate their electronic and magnetic properties and to identify promising candidates for realizing the QAHE.

\begin{table}[!t]
%\centering
%\begin{threeparttable}
\caption{Overview of previous works based on DFT calculation for transition-metal trihalide monolayers. %See footnote  for details \tablefootnote[2]{ In this work, \textit{Dirac half-metal} (DHM) is defined as a material in which one spin channel exhibits a Dirac-like linear crossing at the Fermi level, while the opposite spin channel has a finite band gap and is insulating. \textit{QAHI} denotes a quantum anomalous Hall insulator, i.e., a gapped system with a finite Chern number.  \textit{Ins.} refers to a trivial (non-topological) insulator.  In contrast, \textit{SM} (semimetal) is defined as a system where valence and conduction bands overlap slightly near the Fermi level, with both spin channels contributing to electronic conduction. }.
%\red{((How to make the footnote as reference number?))}
\label{table:overview}}
\setlength{\tabcolsep}{7pt}
\renewcommand{\arraystretch}{1.0}
\begin{tabular}{l c c c c c} 
\hline
\hline
MX$_3$ & Ins. & SM & DHM & QAHI\\ 
\hline
VF$_3$ &\cite{Tomar2019} & – & – & – \\
VCl$_3$ & \cite{Tomar2019}\cite{Mastrippolito2023} & – & \cite{He2016} & –\\
VBr$_3$ & \cite{Tomar2019} & – & – & – \\
VI$_3$ & \cite{Nguyen2021}\cite{Tomar2019} & – & \cite{He2016} & – \\
\\
CrF$_3$   & \cite{Tomar2019} & – & – & – \\
CrCl$_3$  & \cite{Tomar2019} & – & – & – \\
CrBr$_3$  & \cite{Tomar2019} & – & – & – \\
CrI$_3$   & \cite{Nguyen2021}\cite{Tomar2019}& – & – & – \\
\\
MnF$_3$   & – & – & \cite{Tomar2019} & – \\
MnCl$_3$  & – & – & \cite{Tomar2019} & – \\
MnBr$_3$  & – & – & \cite{Tomar2019} & \cite{QSun2018} \\
MnI$_3$   & – & – & \cite{Tomar2019} & – \\
\\
FeF$_3$   & \cite{Tomar2019} & – & – & – \\
FeCl$_3$  & \cite{Tomar2019} & – & \cite{ZhengLi2019} & \cite{PingLi2019}\cite{Olsen2019} \\
FeBr$_3$  & \cite{Tomar2019} & – & \cite{ZhengLi2019} & \cite{PingLi2019}\cite{Olsen2019} \\
FeI$_3$   & \cite{Tomar2019} & – & \cite{ZhengLi2019} & \cite{PingLi2019} \\
\\
NiF$_3$   & – & \cite{Tomar2019} & – & – \\
NiCl$_3$  & – & \cite{Tomar2019} & \cite{ZhengLi2019} & \cite{JHe2017_nicl3} \\
NiBr$_3$  & – & \cite{Tomar2019} & \cite{ZhengLi2019} & – \\
NiI$_3$   & – & \cite{Tomar2019} & \cite{ZhengLi2019} & – \\
\\
PdF$_3$   & – &  – & – &  – \\
PdCl$_3$  & – & \cite{Bafekry2020} & – & \cite{Bafekry2020, Wang2018} \\
PdBr$_3$  & – & \cite{You2019} & – & \cite{You2019} \\
PdI$_3$   & – & – & – & \cite{Olsen2019} \\
\hline
\hline
\end{tabular}
%\end{threeparttable}
\end{table}

\begin{table}[!b]
%\centering
\caption{
Calculated in-plane lattice constant, energy difference between AFM and FM states ($\Delta E = E_{\rm AFM}-E_{\rm FM}$), transition-metal spin moment ($S_M$) and band gap energy with SOC ($E_{\rm gap}$) and Chern number for $MX_3$ monolayer with GGA+$U$ functional. For reference, the nominal $d$-orbital occupations for high-spin $M^{3+}$ ions are:  V: $t_{2g}^2 e_g^0$, Cr: $t_{2g}^3 e_g^0$, Mn: $t_{2g}^3 e_g^1$, Fe: $t_{2g}^3 e_g^2$, Ni: $t_{2g}^6 e_g^1$, and Pd: $t_{2g}^6 e_g^1$.
}
\label{table:mx3_GGA+U}
\setlength{\tabcolsep}{6pt}
\renewcommand{\arraystretch}{1.0}
  \begin{tabular}{l c c c c c} 
  \hline
  \hline
  MX$_3$ & a & $\Delta E$  & $S_{M}$  & $E_{\rm gap}$ & Chern   \\ 
    & ({\AA}) &  (meV) &  ($\mu_{\rm B}$) & (eV) & number  \\ 
  \hline
VF$_3$ & 5.32  & 23.17 & 1.95 & 1.23 & 0\\
VCl$_3$ & 6.16  & 29.23 & 2.01 & 0.88 & 0\\
VBr$_3$ & 6.54  & 28.91 & 2.08 & 0.72 & 0\\
VI$_3$ & 7.10  & 109.37 & 2.43 & 0.57 & 0\\
\\
CrF$_3$ & 5.23 & 19.53 & 2.94 & 2.55 & 0\\
CrCl$_3$ & 6.10  & 34.81 & 3.08 & 2.14 & 0\\
CrBr$_3$ & 6.48  & 45.33 & 3.20 & 1.74 & 0\\
CrI$_3$ & 7.06  & 58.06 & 3.39 & 1.09 & 0\\
\\
\textbf{MnF$_3$} & 5.38  & 506.53 & 3.70 & 0.0185 &  +1\\
\textbf{MnCl$_3$} & 6.25  & 140.57 & 3.98  & 0.0131 &  $+1$ \\
\textbf{MnBr$_3$ }& 6.61  & 126.23 & 4.08 & 0.0345 & $-1$\\
MnI$_3$ & 7.08  & 106.10 & 4.22 & - & -\\
\\
FeF$_3$ & 5.32  & -61.13 & 4.25 & 2.19 & 0\\
FeCl$_3$ & 6.20  & -17.82 & 3.80 & 1.27 & 0\\
FeBr$_3$ & 6.58  & -3.83 & 3.89& 1.01 & 0\\
FeI$_3$ & 7.15  & -4.38  & 3.75 & 0.72 & 0\\
\\
NiF$_3$ & 5.08  & 295.29 & 1.03 & -%0.0002
& - \\
NiCl$_3$ & 5.94  & 147.61 & 0.92 & - & -\\
NiBr$_3$ & 6.30  & 145.84 & 0.95  & - & -\\
NiI$_3$ & 6.75  & 146.08 & 0.98 & - & -\\
\\
\textbf{PdF$_3$} & 5.58  & 416.97 & 0.77 & 0.1277 & $-1$\\
\textbf{PdCl$_3$} & 6.32  & 199.42 & 0.59 & 0.0071 & +1 \\
\textbf{PdBr$_3$} & 6.61  & 170.03 & 0.53 & 0.0439 & $-1$ \\
\textbf{PdI$_3$} & 7.11  & 159.53 & 0.39 & 0.0324 & $-1$\\ 
\hline
\hline
\end{tabular}
\end{table}

\section{Computational Methods}
Density functional theory (DFT) calculations were performed using the VASP package \cite{vasp}. The exchange–correlation functional was treated within the Perdew–Burke–Ernzerhof generalized gradient approximation with an effective Hubbard $U$ correction (GGA+$U$, $U=2$ eV), applied to the $d$ orbitals of transition-metal atoms \cite{gga, Liechtenstein, Nguyen2021}. The projector augmented-wave (PAW) method was used for pseudopotentials \cite{paw}. %Energy cutoffs were set to 65 Ry for the plane-wave basis and 780 Ry for the charge density. 
Monolayer slabs were separated by 15 Å of vacuum to avoid spurious interactions. The lattice constants were optimized in the ferromagnetic configuration (Table \ref{table:mx3_GGA+U}).

Structural relaxations used a $6\times6\times1$ $k$-point mesh until forces were below $1\times10^{-4}$ eV/Å. SOC was included self-consistently, with the spin quantization axis along the hexagonal [0001] axis. Densities of states and total energies were evaluated with a $12\times12\times1$ $k$-point mesh. Topological properties were obtained via maximally localized Wannier functions constructed with WANNIER90 \cite{wannier90} and analyzed using WannierTools \cite{wanniertools}. Chemical bonding analyses were conducted using LOBSTER \cite{LOBSTER}, followed by post-processing with VASPKIT \cite{VASPKIT}.
\begin{figure}[!tbp]
%\begin{center}
%\vspace{-0.3cm}
%\resizebox{120mm}{!}
{
 \includegraphics[width=80mm, angle=0]{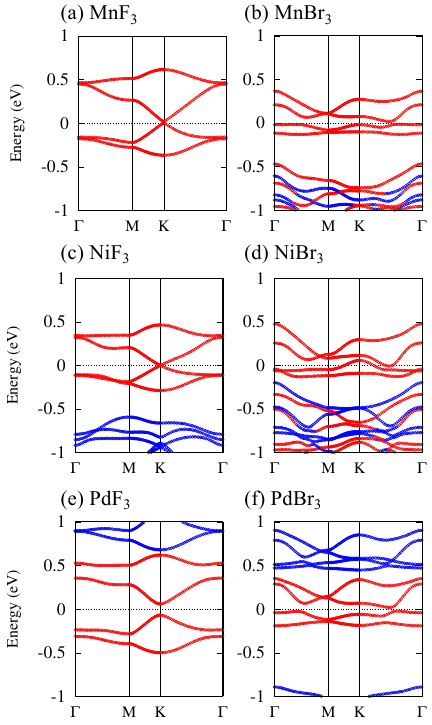}
%\vspace {-0.4cm}
}
\caption{\label{fig:gga+soc}
Bandstructure obtained for monolayer with SOC taken into account in GGA+$U$ calculation. The spin polarization along the $z$ direction ($S_z$) is shown by red (up spin) and blue (down spin) colors. The Fermi energy $E_{\rm F}$ is set as origin of energy. %\red{((Suggested by KY: Plot the bandstructure (with SOC) for VF3, CrF3, FeF3 together with the schematic spin configuration like Fig. 5))}%{\blue{ Change the allignment to 2x3 (not 3x2) to make the band larger, shift the EF of PdF$_3$ to not locate at the top valance band.}}
%\ky{Set red/blue color for spin-polarization of bands. Put the labels like (a) MnF3 out of the figure.}
}
%\vspace {-1.0cm}
%\end{center}
\end{figure} 

\begin{figure*}[!ht]
%\begin{center}
%\vspace{-0.3cm}
%\resizebox{120mm}{!}
{
 \includegraphics[width=140mm, angle=0]{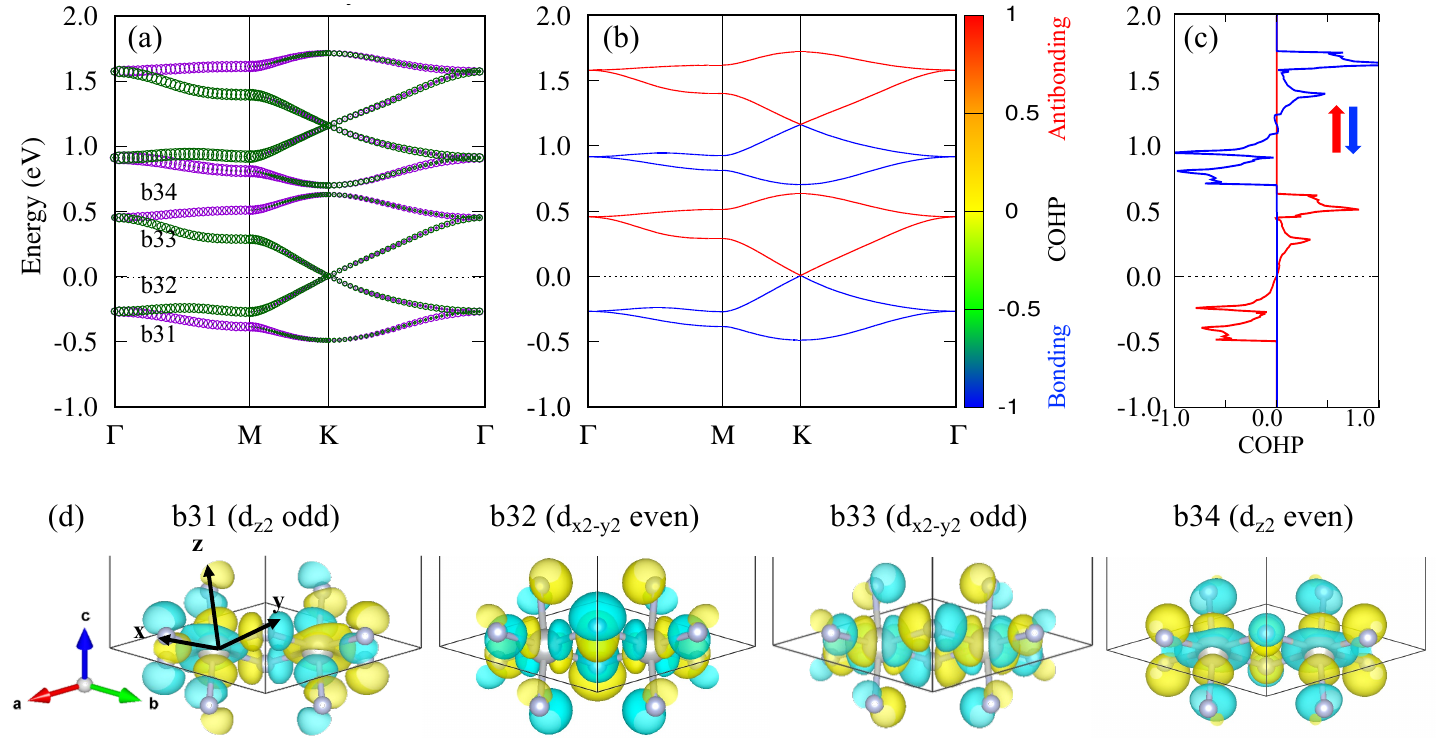}
%\vspace {-0.4cm}
}
\caption{\label{fig:cohp}
(a) Orbital projection of Pd 3$d$ contributions to the bandstructure in PdF$_3$ without SOC. Purple (green) color band presents $d_{z^2}$ ($d_{x^2-y^2}$) orbital projection.
(b) $k$-dependent crystal
orbital Hamilton population (COHP) of Pd 3$d$ $e_g$ state. (c) COHP diagrams for Pd-Pd neighboring atoms. The Fermi level is set as the energy reference at zero. The COHP curves of the up-spin (down-spin) state are shown in red (blue). The negative regions represent the bonding interactions, while the positive regions denote antibonding interactions. (d) The wavefunctions (real part) at the $\Gamma$ point corresponding to 31st to 34th bands (b31-b34) with odd/even parity. The positive and negative sign is shown by yellow and cyan colors \cite{VASPKIT}. %The isosurface level was set at 1.87$\times 10^{-7}$ $a_0^{−3/2}$ \blue{(??)} \red{(TN took this value from VESTA) where $a_0$ is the Bohr radius.}
 %Isosurface 1.86789e-07
}
%\vspace {-1.0cm}
%\end{center}
\end{figure*} 

\section{Results}
\subsection{Electronic and magnetic properties}
The calculated in-plane lattice constants, energy differences between antiferromagnetic (AFM) and ferromagnetic (FM) configurations, transition-metal spin moments, and band gaps with SOC of the $MX_3$ monolayers %using GGA+$U$ ($U$ = 2.0 eV) potential 
 are summarized in Table  \ref{table:mx3_GGA+U} 
\footnote[2]{ %\blue{((do you need it?))}
 Note that, the in-plane lattice constants are carefully optimized to reach the equilibrium state within the GGA and GGA+$U$ framework. %[see Table \ref{table:mx3_bareGGA} and Table \ref{table:mx3_GGA+U} and Supplemental Material]. 
The difference in lattice constant value between GGA and GGA+$U$ approach is negligibly small. %However, in the case of VX$_3$, as the authors pointed out in the previous work of VI$_3$, lattice constant in metallic state (5.39{\AA}, 6.25{\AA}, 6.64{\AA} and 7.16{\AA} for VF$_3$, VCl$_3$, VBr$_3$ and VI$_3$, respectively ) is predicted larger than the one in insulating state (5.32{\AA}, 6.16{\AA}, 6.54{\AA} and 7.10{\AA}) driven by the on-site Coulomb repulsion.  
}.
Overall, the data reveal clear trends %in both magnetic ordering and electronic structure 
 across the family of transition-metal trihalides. 
%In particular, the compounds VX$_3$, CrX$_3$ and FeX$_3$ consistently exhibit positive band gaps, indicating insulating behavior. In which, only FeX$_3$ shows negative E$_{\rm AFM-FM}$ values indicating their ground state of AFM. This contrasts with most other compounds, where they have positive E$_{\rm AFM-FM}$ values indicating FM ground state, as seen in VX$_3$ and CrX$_3$. Interestingly, MnX$_3$ and NiX$_3$ and PdF$_3$ compounds exhibit zero band gaps despite strong ferromagnetism, suggesting half-metallic or semimetallic behavior, which can support spin-polarized conduction.
VX$_3$ and CrX$_3$ compounds are found to be ferromagnetic insulators, whereas FeX$_3$ exhibits antiferromagnetic insulating behavior, consistent with experimental reports for FeCl$_3$\cite{Cable1962,KOCHER1967}.  The calculated band structures for monolayer VF$_3$, CrF$_3$, and FeF$_3$ indicate that these materials exhibit trivial insulating behavior.
In contrast, MnX$_3$, NiX$_3$, and PdF$_3$ show strong ferromagnetism with vanishing or very small band gaps, indicative of half-metallic or semimetallic states that may support spin-polarized transport. 

In VI$_3$, the electronic structure is highly sensitive to the Hubbard $U$ correction; with $U=2$ eV a ferromagnetic Mott insulating phase is stabilized, consistent with optical spectroscopy measurements and our earlier work\cite{Nguyen2021, vi3_cri3_arpes, Kong2019}.

In Fe$X_3$, the high-spin configuration of Fe$^{3+}$ is energetically favored, stabilizing an AFM ground state in agreement with neutron diffraction and Mössbauer data \cite{Cable1962,KOCHER1967}.

%\blue{(KY comes here 20250820)}

As shown in Fig. \ref{fig:gga+soc},  
MnF$_3$, NiF$_3$, and PdF$_3$  exhibit the fully spin-polarized Dirac cone at the $K$ point 
manifesting the Dirac half metallicity. SOC makes these Dirac-cone bands gapped, which band-gap size depends on the strength of SOC in either the metal or halogen atoms. Especially, PdF$_3$ shows the sizable band gap ($E_{\rm gap} > $ 0.1 eV) due to the strong SOC of the Pd 4$d$ orbital state, providing the suitable condition for demonstrating the QAHE. 
However, upon replacing fluorine with the heavier bromine ligand [Figs.  \ref{fig:gga+soc}(b, d, f)], the enhanced SOC associated with Br atoms leads to a reconstruction of the band topology. The Dirac cone at the $K$ point becomes distorted or fully gapped, reflecting strong SOC-induced hybridization between the spin-up and spin-down channels. %This progression from MF$_3$ to MBr$_3$ 
This behavior highlights the crucial role of halogen substitution in tuning the SOC strength and, consequently, the topological characteristics of the M$X_3$ monolayers.
%This behavior underscores that chemical substitution of the halogen element effectively tunes the SOC strength and thereby dictates the emergence or suppression of topological electronic states in the MX$_3$ monolayers.

\subsection{Orbital character and chemical bonding in PdF$_3$}

Before addressing the topological properties, we analyze the orbital character and bonding interactions in PdF$_3$. Fig. \ref{fig:cohp}(a) shows that the low-energy states are mainly derived from Pd-$e_g$ orbital state, consistent with $d^7$ ($t_{2g}^6e_g^1$) configuration. The Dirac crossing at the $K$ point originates from alternating $d_{z^2}$ and $d_{x^2-y^2}$ components on the honeycomb lattice, forming the $\sigma$ bonds. 

The COHP analysis [Figs. \ref{fig:cohp}(b, c)] clarifies that electronic states below the Fermi level have predominantly bonding character, while those above have antibonding character, with the Dirac point marking the transition between the two. Direct Pd–Pd $\sigma$ interactions dominate this dispersion, while Pd–F contributions are comparatively weaker. This picture supports a simple two-orbital model, where bonding–antibonding splitting defines the Dirac cone, 
analogous to graphene in which the Dirac cone arises from the bonding and antibonding combination of C–$p_z$ orbitals on the honeycomb-pattern lattice \cite{CastroNeto2009}. 

A notable feature is that the SOC-induced gap at the Dirac point remains modest even when Hubbard $U$ correction is included. Because occupied and unoccupied $e_g$ states hybridize only weakly across the Fermi level through SOC, the effect of $U$ on the band dispersion is limited. Thus, unlike other correlated 2D magnets where $U$ strongly enhances the gap, in PdF$_3$ electronic correlations play only a secondary role. %\red{((Add U-dependence bandstructure figure in SI ))}

%\red{TN: write disscussion of FIG 2!}

 \subsection{Quantum anomalous Hall effect}

The anomalous Hall conductivity (AHC) was calculated as integrating the Berry curvatures with a summation over the occupied states in the Brillouin zone by the WANNIER90 code \cite{wannier90}. In a two-dimensional system, the AHC is given by the integration of the Berry curvature over the Brillouin zone (BZ) for all occupied bands:

\begin{equation}
\sigma_{xy} = -\frac{e^2}{\hbar} \sum_{n \in \text{occ}} \int_{\text{BZ}} \frac{d^2\mathbf{k}}{(2\pi)^2} \, \Omega_n^z(\mathbf{k}),
\end{equation}

where \( \Omega_{n,xy}(\mathbf{k}) \) is the Berry curvature of the \( n \)-th band at momentum \( \mathbf{k} \), and \( e \) and \( \hbar \) are the elementary charge and the reduced Planck constant, respectively.

The Berry curvature was calculated  %written as an antisymmetric tensor
by 
%\thao{Eq. below taken from user_guide_wannier90_ver3.1}

\begin{eqnarray}
    \Omega^{z}_{n}({\bm k})=-2{\rm Im}\left< \nabla_{k_x}u_{n{\bm k}}|\nabla_{k_y}u_{n{\bm k}} \right> 
\end{eqnarray} 
in Wannier90 code\cite{LVTS12, wannier90}. 
%The derivative $\nabla_{k_\alpha}u_{n{\bm k}}$ can be obtained 
Here, $u_{n{\bm k}}$ are the cell-periodic Bloch functions for $n$-th band, projected onto Wannier functions $\left| {\bm R} n \right>$ by 
\begin{eqnarray}
u_{n{\bm k}} = \sum_{\bm R} e^{-i{\bm k}\cdot({\bm r}-{\bm R})}\left| {\bm R} n \right>. 
\end{eqnarray} 

%The Berry curvature \( \Omega_n^z(\mathbf{k}) \) can be expressed by Kubo formula:
%\begin{equation}
%\Omega_{n,xy}(\mathbf{k}) = -2 \, \mathrm{Im} \sum_{m \ne n} \frac{ \langle u_{n\mathbf{k}} | \hat{v}_x | u_{m\mathbf{k}} \rangle \langle u_{m\mathbf{k}} | \hat{v}_y | u_{n\mathbf{k}} \rangle }{ \left( E_{m\mathbf{k}} - E_{n\mathbf{k}} \right)^2 },
%\end{equation}
%where \( |u_{n\mathbf{k}}\rangle \) are the cell-periodic Bloch functions, \( \hat{v}_x \) and \( \hat{v}_y \) are the velocity operators in the \( x \) and \( y \) directions, and \( E_{n\mathbf{k}} \) is the energy of the \( n \)-th band at \( \mathbf{k} \).

%as long as the Fermi level lies within the SOC-induced energy gap. %In the case of PdF\textsubscript{3}, the Berry curvature exhibits pronounced peaks near the K and K′ points in the Brillouin zone, and integration yields \( C = +1 \), confirming the presence of a QAH insulating state with topologically protected edge channels.

\begin{figure}[!hb]
%\begin{center}
%\vspace{-0.3cm}
%\resizebox{120mm}{!}
{
 \includegraphics[width=65mm, angle=0]{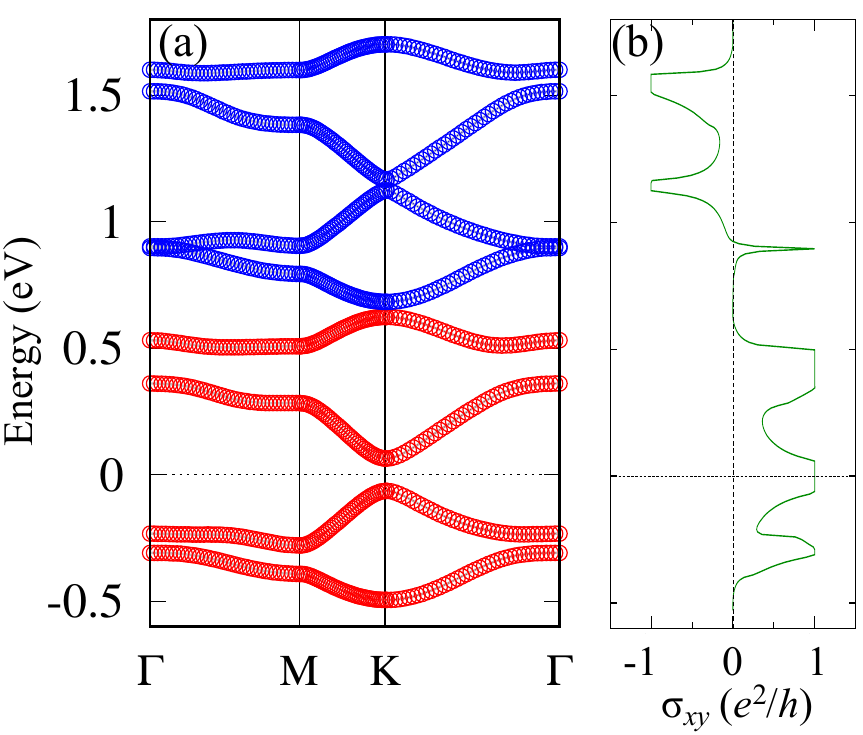}
%\vspace {-0.4cm}
}
\caption{\label{fig:PdF3_AHC}
%\red{((TN:Do we need to switch the sign of sigmaxy to match with Chern number?))}
(a) The bandstructure of monolayer PdF$_3$ with $s_z$ polarization in color, red for up spin and blue for down spin. Fermi energy is set to zero. (b) The calculated anomalous Hall conductivity of monolayer PdF$_3$ as a function of energy near the Fermi level.  %\red{ADD Hybrid WCCs here because it is bulk properties}
%\blue{Replace it by GGA+U result}
%\red{((shift the Fermi energy to the midle of the gap))}
}
%\vspace {-1.0cm}
%\end{center}
\end{figure}

A non-zero integer value of Chern number \( C \) defines a nontrivial topological phase associated with the quantized Hall conductivity, 
\begin{equation}
\sigma_{xy} = -\, \frac{e^2}{h}C.
\end{equation}
%For example, if \( C = \pm 1 \), the system supports a quantized Hall conductivity:
As shown in Fig. \ref{fig:PdF3_AHC}(a), % highlights the topological and magnetic properties of monolayer PdF$_3$, presenting both the spin-resolved bandstructure projected onto the z-component of spin $s_z$ and the corresponding anomalous Hall conductivity (AHC) as a function of energy. T
the fully spin-polarized Dirac cone at the K point becomes gapped by the present of SOC %. This gap opens selectively in one spin channel, 
confirming that PdF$_3$ is QAHI,  % with broken time-reversal symmetry and intrinsic spin polarization. 
%The presence of a spin-resolved, SOC-induced band gap strongly indicates topologically nontrivial behavior. 
supported by the calculated AHC (see Fig. \ref{fig:PdF3_AHC}(b)) being a quantized number consistent with the Chern number, $C=-1$, in the SOC-induced band gap. %In particular, the peak in the AHC appears sharply centered around the Fermi energy (E = 0 eV), precisely at the SOC-induced gap. 
%This alignment reinforces that the topological gap is responsible for the quantized Hall response and further confirms the QAHE phase. %The sharply peaked Berry curvature around the $K$ point contributes dominantly to the AHC, underscoring the role of valley-resolved topology in the bandstructure. These findings position PdF$_3$ as a rare example of a stoichiometric, intrinsic 2D QAHE material, offering a realistic path toward experimental realization of dissipationless edge conduction in spintronic devices. Additionally, the use of a light halogen (F) in PdF$_3$—unlike heavier halides traditionally associated with strong SOC—suggests that the transition metal center (i.e. Pd) dominates the topological characteristics, offering new chemical degrees of freedom for tuning QAHE in related systems.

 \begin{figure*}[htbp]
%\begin{center}
%\vspace{-0.3cm}
%\resizebox{120mm}{!}
{
 \includegraphics[width=150mm, angle=0]{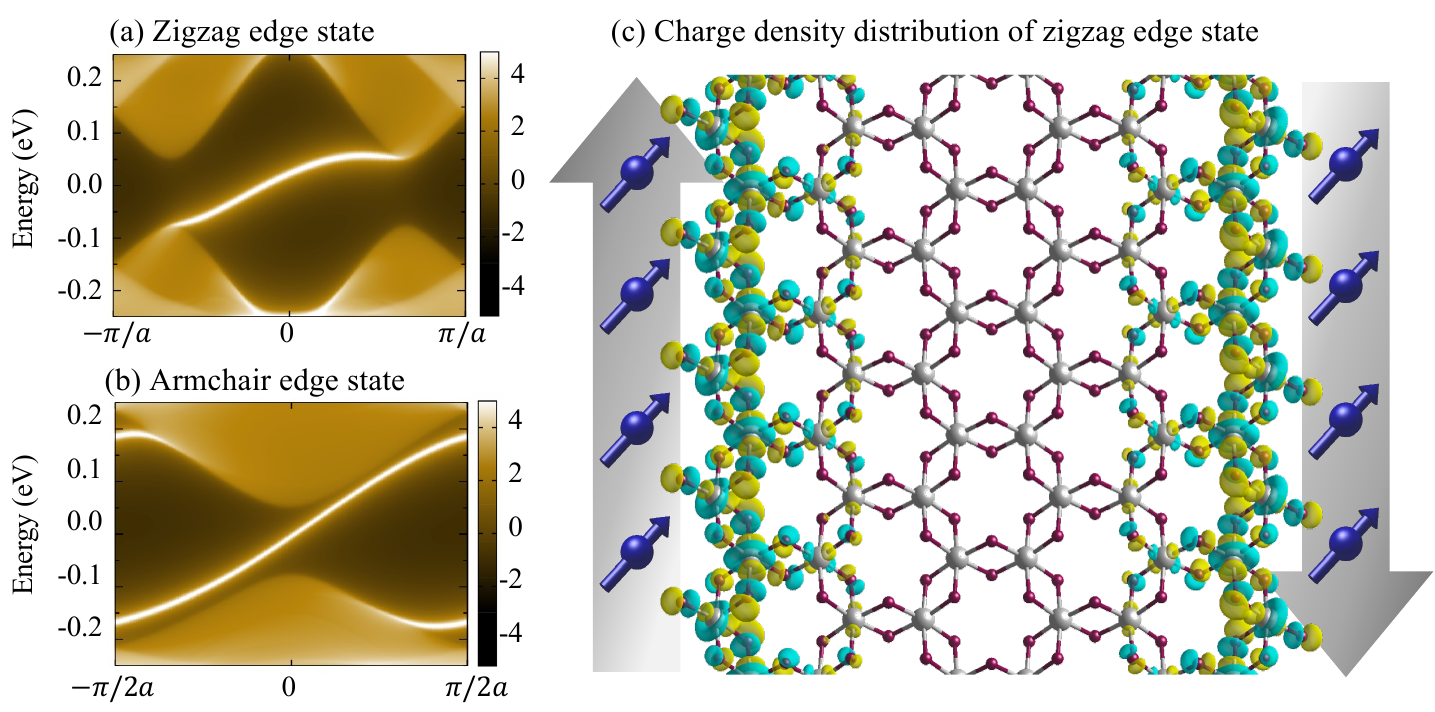}
%\vspace {-0.4cm}
}
\caption{\label{fig:pdF3_edge}
%\blue{UPDATE with U=2.0eV result. Correct the range in Fig.b}
Bandstructure of (a) the zigzag and (b) the armchair edges of the PdF$_3$ nanoribbon, with the edge states connecting
the 2D valence and conduction bands. (c) The charge density distribution in the zigzag edge of PdF$_3$ nanoribbon. %\red{(TN: check the direction of the spin flow)}
}
%\vspace {-1.0cm}
%\end{center}
\end{figure*}

The bulk–edge correspondence in QAHI guarantees the existence of chiral edge states.
To verify this explicitly, we performed slab-model band calculations for PdF$_3$ nanoribbons with zigzag and armchair terminations, as shown in Figs. \ref{fig:pdF3_edge}(a,b). In both cases, a single edge band traverses the bulk band gap and connects the valence and conduction bands near the Fermi level
%in the vicinity of the Fermi level in the zigzag edge termination. 
The corresponding states are exponentially localized at the ribbon boundaries, as confirmed by edge-projected spectral weight and the charge-density plot shown in Fig. \ref{fig:pdF3_edge} (c). 
Their group velocity retains a fixed sign along a given edge, characteristic of a chiral mode that is robust against elastic backscattering. In particular, up-spin electrons propagate unidirectionally along one boundary of the ribbon, while the opposite edge hosts electrons moving in the counter direction.
These gap-crossing edge states represent topologically protected boundary modes in the QAH phase and are fully consistent with the nonzero Chern number obtained from our bulk Wannier-based analysis.
%confirming that PdF$_3$ is a Chern insulator with nontrivial topology.

%The asymmetry and slope of these states indicate unidirectional, dissipationless edge transport—a hallmark of QAHE.

%To further substantiate the topological nature of the system, the authors employ the hybrid Wannier charge centers (WCC) method, as shown in FIg. 4(c). The hybird WCCs, computed via a maximally localized Wannier function approach, evolve continuously across the Brillouin zone and clearly wind across the unit cell—producing a nonzero Chern number (i.e. C = +1). This indicates that one quantum of charge is pumped across the unit cell per cycle, consistent with quantized anomalous Hall conductivity  and in full agreement with the results from Fig. 3.

%The combination of the bulk bandstructure (Dirac cone with SOC gap), quantized anomalous Hall conductivity, and gapless edge states in the zigzag and armchair directions forms a comprehensive topological characterization of PdF$_3$ as a robust intrinsic 2D QAHE material. Unlike magnetically doped topological insulators that require ultralow temperatures, the strong intrinsic ferromagnetism and sizable topological gap in PdF$_3$ make it a promising candidate for high-temperature QAH applications, rivaling or even surpassing other proposed QAH systems such as MnBr$_3$. %This figure therefore solidifies PdF’s potential for practical spintronic devices based on topologically protected edge conduction.
 
%\section{Discussion}

\section{Conclusion}
By means of density-functional-theory calculations, we investigated the QAHE in monolayer $MX_3$ ($M$ = V, Cr, Mn, Fe, Ni, Pd; $X$ = F, Cl, Br, I). Among them, V$X_3$, Cr$X_3$ and Fe$X_3$ are insulating, while Ni$X_3$, Mn$X_3$, and Pd$X_3$ exhibit semimetallic character.  

For PdF$_3$, the band structure shows a fully spin-polarized Dirac cone at the $K$ point. The inclusion of SOC lifts the degeneracy and opens a sizable gap, yielding a nonzero Chern number of one as determined from Wannier-function calculations. The emergence of gap-crossing chiral edge states in nanoribbon geometries further confirms its topologically nontrivial character.  

Taken together, the Dirac cone with an SOC-induced gap, the quantized anomalous Hall conductivity, and the gapless edge modes establish PdF$_3$ as an intrinsic 2D QAHE material. Unlike magnetically doped topological insulators that require ultralow temperatures, PdF$_3$ combines strong intrinsic ferromagnetism with a sizable topological gap, making it a promising candidate for high-temperature QAHE applications. Beyond fundamental interest, such materials may also open avenues toward low-power spintronic devices and fault-tolerant quantum information technologies based on topologically protected states.

%\blue{---KY has come here, and will lead this passage to the very end (as curved in a poneglyph 20250827). }

\begin{comment}\blue{Below is moved from Result. but do we need to write such a unreliable applications??}
The presence of SOC-induced gaps at symmetry-protected Dirac points in a single spin channel confirms the nontrivial topological character of these materials. Furthermore, the preservation of spin polarization across the band gap hints at potential applications in spin-filtering or spin-injection devices, where one spin state dominates conduction. The contrast between different compounds in this figure also emphasizes the role of chemical composition in tuning electronic topology, and suggests that even subtle changes in halogen choice can switch a material from trivial to topological. This highlights the promise of MX$_3$ monolayers as a versatile platform for spintronic applications driven by intrinsic magnetic ordering and band topology.
\end{comment}
\section*{Acknowledgments}
This work was supported by %\blue{Add Sato CREST number and Tabata CREST number}.
%Sato CREST 
JST-CREST (Grant No. JPMJCR18T1) and 
%Tabata CREST
JST-CREST (Grant No. JPMJCR22O2) and by Institute for Open and Transdisciplinary Research Initiatives (OTRI), Osaka University.
 The numerical computation was performed on the Supercomputing Facilities at the Institute for Solid State Physics, University of Tokyo and the Flow Cloud system at Nagoya University . %The authors also acknowledge Center for Computational Materials Science, Institute for Materials Research, Tohoku University for the use of MASAMUNE-IMR (Project No.202312-RDKGE-0058). 

%\section*{Appendix}
%\appendix 

\section*{References}
\nocite{*}
\bibliographystyle{apsrev4-1}
\bibliography{mybibitem.bib}

@PREAMBLE{
 "\providecommand{\noopsort}[1]{}" 
 # "\providecommand{\singleletter}[1]{#1}%" 
}

@article{CastroNeto2009,
  title = {The electronic properties of graphene},
  author = {Castro Neto, A. H. and Guinea, F. and Peres, N. M. R. and Novoselov, K. S. and Geim, A. K.},
  journal = {Rev. Mod. Phys.},
  volume = {81},
  issue = {1},
  pages = {109--162},
  numpages = {0},
  year = {2009},
  month = {Jan},
  publisher = {American Physical Society},
  doi = {10.1103/RevModPhys.81.109},
  url = {https://link.aps.org/doi/10.1103/RevModPhys.81.109}
}

@Article{He2016,
author ="He, Junjie and Ma, Shuangying and Lyu, Pengbo and Nachtigall, Petr",
title  ="Unusual Dirac half-metallicity with intrinsic ferromagnetism in vanadium trihalide monolayers",
journal  ="J. Mater. Chem. C",
year  ="2016",
volume  ="4",
issue  ="13",
pages  ="2518-2526",
publisher  ="The Royal Society of Chemistry",
doi  ="10.1039/C6TC00409A",
url  ="http://dx.doi.org/10.1039/C6TC00409A",
}

@article{Tomar2019,
title = {Intrinsic magnetism in monolayer transition metal trihalides: A comparative study},
journal = {Journal of Magnetism and Magnetic Materials},
volume = {489},
pages = {165384},
year = {2019},
issn = {0304-8853},
doi = {https://doi.org/10.1016/j.jmmm.2019.165384},
url = {https://www.sciencedirect.com/science/article/pii/S0304885319301301},
author = {Shalini Tomar and Barun Ghosh and Sougata Mardanya and Priyank Rastogi and B.S. Bhadoria and Yogesh Singh Chauhan and Amit Agarwal and Somnath Bhowmick}
}

@article{QSun2018,
  title = {Prediction of manganese trihalides as two-dimensional Dirac half-metals},
  author = {Sun, Qilong and Kioussis, Nicholas},
  journal = {Phys. Rev. B},
  volume = {97},
  issue = {9},
  pages = {094408},
  numpages = {5},
  year = {2018},
  month = {Mar},
  publisher = {American Physical Society},
  doi = {10.1103/PhysRevB.97.094408},
  url = {https://link.aps.org/doi/10.1103/PhysRevB.97.094408}
}

@Article{Wang2018,
author ="Wang, Ya-ping and Li, Sheng-shi and Zhang, Chang-wen and Zhang, Shu-feng and Ji, Wei-xiao and Li, Ping and Wang, Pei-ji",
title  ="High-temperature Dirac half-metal PdCl3: a promising candidate for realizing quantum anomalous Hall effect",
journal  ="J. Mater. Chem. C",
year  ="2018",
volume  ="6",
issue  ="38",
pages  ="10284-10291",
publisher  ="The Royal Society of Chemistry",
doi  ="10.1039/C8TC02500B",
url  ="http://dx.doi.org/10.1039/C8TC02500B"
}

@article{Bafekry2020,
	author = {Bafekry, Asadollah and Stampfl, Catherine and Peeters, Francois M.},
	da = {2020/01/14},
	doi = {10.1038/s41598-019-57353-3},
	id = {Bafekry2020},
	isbn = {2045-2322},
	journal = {Scientific Reports},
	number = {1},
	pages = {213},
	title = {Dirac half-metallicity of Thin PdCl3 Nanosheets: Investigation of the Effects of External Fields, Surface Adsorption and Defect Engineering on the Electronic and Magnetic Properties},
	ty = {JOUR},
	url = {https://doi.org/10.1038/s41598-019-57353-3},
	volume = {10},
	year = {2020}}

@article{Olsen2019,
  title = {Discovering two-dimensional topological insulators from high-throughput computations},
  author = {Olsen, Thomas and Andersen, Erik and Okugawa, Takuya and Torelli, Daniele and Deilmann, Thorsten and Thygesen, Kristian S.},
  journal = {Phys. Rev. Materials},
  volume = {3},
  issue = {2},
  pages = {024005},
  numpages = {11},
  year = {2019},
  month = {Feb},
  publisher = {American Physical Society},
  doi = {10.1103/PhysRevMaterials.3.024005},
  url = {https://link.aps.org/doi/10.1103/PhysRevMaterials.3.024005}
}

@Article{PingLi2019,
author ="Li, Ping",
title  ="Prediction of intrinsic two dimensional ferromagnetism realized quantum anomalous Hall effect",
journal  ="Phys. Chem. Chem. Phys.",
year  ="2019",
volume  ="21",
issue  ="12",
pages  ="6712-6717",
publisher  ="The Royal Society of Chemistry",
doi  ="10.1039/C8CP07781A",
url  ="http://dx.doi.org/10.1039/C8CP07781A",
}

@article{Zhou2016,
	author = {Zhou, Yungang and Lu, Haifeng and Zu, Xiaotao and Gao, Fei},
	da = {2016/01/18},
	doi = {10.1038/srep19407},
	id = {Zhou2016},
	isbn = {2045-2322},
	journal = {Scientific Reports},
	number = {1},
	pages = {19407},
	title = {Evidencing the existence of exciting half-metallicity in two-dimensional TiCl3 and VCl3 sheets},
	ty = {JOUR},
	url = {https://doi.org/10.1038/srep19407},
	volume = {6},
	year = {2016}}

@article{Nguyen2021,
  title = {Electric-field tuning of the magnetic properties of bilayer ${\mathrm{VI}}_{3}$: A first-principles study},
  author = {Nguyen, Thi Phuong Thao  and Yamauchi, Kunihiko and Oguchi, Tamio and Amoroso, Danila and Picozzi, Silvia},
  journal = {Phys. Rev. B},
  volume = {104},
  issue = {1},
  pages = {014414},
  numpages = {8},
  year = {2021},
  month = {Jul},
  publisher = {American Physical Society},
  doi = {10.1103/PhysRevB.104.014414},
  url = {https://link.aps.org/doi/10.1103/PhysRevB.104.014414}
}

@article{You2019,
  title = {Two-Dimensional Room-Temperature Ferromagnetic Semiconductors with Quantum Anomalous Hall Effect},
  author = {You, Jing-Yang and Zhang, Zhen and Gu, Bo and Su, Gang},
  journal = {Phys. Rev. Applied},
  volume = {12},
  issue = {2},
  pages = {024063},
  numpages = {7},
  year = {2019},
  month = {Aug},
  publisher = {American Physical Society},
  doi = {10.1103/PhysRevApplied.12.024063},
  url = {https://link.aps.org/doi/10.1103/PhysRevApplied.12.024063}
}

@article{Mastrippolito2023,
  title = {Polaronic and Mott insulating phase of layered magnetic vanadium trihalide ${\mathrm{VCl}}_{3}$},
  author = {Mastrippolito, Dario and Camerano, Luigi and \ifmmode \acute{S}\else \'{S}\fi{}wi\k{a}tek, Hanna and \ifmmode \check{S}\else \v{S}\fi{}m\'{\i}d, B\ifmmode \check{r}\else \v{r}\fi{}etislav and Klimczuk, Tomasz and Ottaviano, Luca and Profeta, Gianni},
  journal = {Phys. Rev. B},
  volume = {108},
  issue = {4},
  pages = {045126},
  numpages = {7},
  year = {2023},
  month = {Jul},
  publisher = {American Physical Society},
  doi = {10.1103/PhysRevB.108.045126},
  url = {https://link.aps.org/doi/10.1103/PhysRevB.108.045126}
}

@Article{JHe2017_nicl3,
author ="He, Junjie and Li, Xiao and Lyu, Pengbo and Nachtigall, Petr",
title  ="Near-room-temperature Chern insulator and Dirac spin-gapless semiconductor: nickel chloride monolayer",
journal  ="Nanoscale",
year  ="2017",
volume  ="9",
issue  ="6",
pages  ="2246-2252",
publisher  ="The Royal Society of Chemistry",
doi  ="10.1039/C6NR08522A",
url  ="http://dx.doi.org/10.1039/C6NR08522A"}

@Article{ZhengLi2019,
author ="Li, Zheng and Zhou, Baozeng and Luan, Chongbiao",
title  ="Strain-tunable magnetic anisotropy in two-dimensional Dirac half-metals: nickel trihalides",
journal  ="RSC Adv.",
year  ="2019",
volume  ="9",
issue  ="61",
pages  ="35614-35623",
publisher  ="The Royal Society of Chemistry",
doi  ="10.1039/C9RA06474E",
url  ="http://dx.doi.org/10.1039/C9RA06474E"}

@ARTICLE{vi3_ml3_insu_soc,
  title = {${\mathrm{VI}}_{3}$: A two-dimensional Ising ferromagnet},
  author = {Yang, Ke and Fan, Fengren and Wang, Hongbo and Khomskii, D. I. and Wu, Hua},
  journal = {Phys. Rev. B},
  volume = {101},
  issue = {10},
  pages = {100402(R)},
  numpages = {5},
  year = {2020},
  month = {Mar},
  publisher = {American Physical Society},
  doi = {10.1103/PhysRevB.101.100402},
  url = {https://link.aps.org/doi/10.1103/PhysRevB.101.100402}
}

@ARTICLE{Kong2019,
author = {Kong, Tai and Stolze, Karoline and Timmons, Erik I. and Tao, Jing and Ni, Danrui and Guo, Shu and Yang, Zo\UTF{00EB} and Prozorov, Ruslan and Cava, Robert J.},
title = {VI3—a New Layered Ferromagnetic Semiconductor},
journal = {Advanced Materials},
volume = {31},
number = {17},
pages = {1808074},
doi = {https://doi.org/10.1002/adma.201808074},
url = {https://onlinelibrary.wiley.com/doi/abs/10.1002/adma.201808074},
year = {2019}
}

@ARTICLE{vi3_ml1_2states,
  title = {Electronic and magnetic properties of van der Waals ferromagnetic semiconductor ${\mathrm{VI}}_{3}$},
  author = {Wang, Yun-Peng and Long, Meng-Qiu},
  journal = {Phys. Rev. B},
  volume = {101},
  issue = {2},
  pages = {024411},
  numpages = {5},
  year = {2020},
  month = {Jan},
  publisher = {American Physical Society},
  doi = {10.1103/PhysRevB.101.024411},
  url = {https://link.aps.org/doi/10.1103/PhysRevB.101.024411}
}

@ARTICLE{vi3_ml2_metallic,
author ="Huang, Chengxi and Wu, Fang and Yu, Shunli and Jena, Puru and Kan, Erjun",
title  ="Discovery of twin orbital-order phases in ferromagnetic semiconducting VI3 monolayer",
journal  ="Phys. Chem. Chem. Phys.",
year  ="2020",
volume  ="22",
issue  ="2",
pages  ="512-517",
publisher  ="The Royal Society of Chemistry",
doi  ="10.1039/C9CP05643B",
url  ="http://dx.doi.org/10.1039/C9CP05643B"}

@ARTICLE{vi3_cri3_arpes,
	author = {Kundu, Asish K. and Liu, Yu and Petrovic, C. and Valla, T.},
	da = {2020/09/24},
	doi = {10.1038/s41598-020-72487-5},
	id = {Kundu2020},
	isbn = {2045-2322},
	journal = {Scientific Reports},
	number = {1},
	pages = {15602},
	title = {Valence band electronic structure of the van der Waals ferromagnetic insulators: VI{\$}{\$}{\_}3{\$}{\$}3and CrI{\$}{\$}{\_}3{\$}{\$}3},
	ty = {JOUR},
	url = {https://doi.org/10.1038/s41598-020-72487-5},
	volume = {10},
	year = {2020}}

@ARTICLE{vi3_stack,
	author = {Long, Chen and Wang, Tao and Jin, Hao and Wang, Hao and Dai, Ying},
	booktitle = {The Journal of Physical Chemistry Letters},
	da = {2020/03/19},
	date = {2020/03/19},
	doi = {10.1021/acs.jpclett.0c00065},
	journal = {The Journal of Physical Chemistry Letters},
	journal1 = {J. Phys. Chem. Lett.},
	m3 = {doi: 10.1021/acs.jpclett.0c00065},
	month = {03},
	number = {6},
	pages = {2158--2164},
	publisher = {American Chemical Society},
	title = {Stacking-Independent Ferromagnetism in Bilayer VI3 with Half-Metallic Characteristic},
	ty = {JOUR},
	url = {https://doi.org/10.1021/acs.jpclett.0c00065},
	volume = {11},
	year = {2020},
	year1 = {2020}}

@article{vi3_ml_insu_dop,
	author = {An, Ming and Zhang, Yang and Chen, Jun and Zhang, Hui-Min and Guo, Yunjun and Dong, Shuai},
	booktitle = {The Journal of Physical Chemistry C},
	da = {2019/12/19},
	date = {2019/12/19},
	doi = {10.1021/acs.jpcc.9b08706},
	isbn = {1932-7447},
	journal = {The Journal of Physical Chemistry C},
	journal1 = {J. Phys. Chem. C},
	m3 = {doi: 10.1021/acs.jpcc.9b08706},
	month = {12},
	number = {50},
	pages = {30545--30550},
	publisher = {American Chemical Society},
	title = {Tuning Magnetism in Layered Magnet VI3: A Theoretical Study},
	ty = {JOUR},
	url = {https://doi.org/10.1021/acs.jpcc.9b08706},
	volume = {123},
	year = {2019},
	year1 = {2019}}

@article{Zhao2021,
author = {Zhao, Shan and Wan, Wenhui and Ge, Yanfeng and Liu, Yong},
title = {Prediction of Chalcogen-Doped VCl3 Monolayers as 2D Ferromagnetic Semiconductors with Enhanced Optical Absorption},
journal = {Annalen der Physik},
volume = {533},
number = {6},
pages = {2100064},
doi = {https://doi.org/10.1002/andp.202100064},
url = {https://onlinelibrary.wiley.com/doi/abs/10.1002/andp.202100064},
year = {2021}
}

@ARTICLE{vasp,
  title = {Efficient iterative schemes for ab initio total-energy calculations using a plane-wave basis set},
  author = {Kresse, G. and Furthm\"uller, J.},
  journal = {Phys. Rev. B},
  volume = {54},
  issue = {16},
  pages = {11169--11186},
  numpages = {0},
  year = {1996},
  month = {Oct},
  publisher = {American Physical Society},
  doi = {10.1103/PhysRevB.54.11169},
  url = {https://link.aps.org/doi/10.1103/PhysRevB.54.11169}
}

@ARTICLE{gga,
  title = {Generalized Gradient Approximation Made Simple},
  author = {Perdew, John P. and Burke, Kieron and Ernzerhof, Matthias},
  journal = {Phys. Rev. Lett.},
  volume = {77},
  issue = {18},
  pages = {3865--3868},
  numpages = {0},
  year = {1996},
  month = {Oct},
  publisher = {American Physical Society},
  doi = {10.1103/PhysRevLett.77.3865},
  url = {https://link.aps.org/doi/10.1103/PhysRevLett.77.3865}
}

@ARTICLE{wannier90,
title = "wannier90: A tool for obtaining maximally-localised Wannier functions",
journal = "Computer Physics Communications",
volume = "178",
number = "9",
pages = "685 - 699",
year = "2008",
issn = "0010-4655",
doi = "https://doi.org/10.1016/j.cpc.2007.11.016",
url = "http://www.sciencedirect.com/science/article/pii/S0010465507004936",
author = "Arash A. Mostofi and Jonathan R. Yates and Young-Su Lee and Ivo Souza and David Vanderbilt and Nicola Marzari",
}

@article{paw,
  title = {Projector augmented-wave method},
  author = {Bl\"ochl, P. E.},
  journal = {Phys. Rev. B},
  volume = {50},
  issue = {24},
  pages = {17953--17979},
  numpages = {0},
  year = {1994},
  month = {Dec},
  publisher = {American Physical Society},
  doi = {10.1103/PhysRevB.50.17953},
  url = {https://link.aps.org/doi/10.1103/PhysRevB.50.17953}
}

@article{wanniertools,
title = {WannierTools: An open-source software package for novel topological materials},
journal = {Computer Physics Communications},
volume = {224},
pages = {405-416},
year = {2018},
issn = {0010-4655},
doi = {https://doi.org/10.1016/j.cpc.2017.09.033},
url = {https://www.sciencedirect.com/science/article/pii/S0010465517303442},
author = {QuanSheng Wu and ShengNan Zhang and Hai-Feng Song and Matthias Troyer and Alexey A. Soluyanov}
}

@article{Liechtenstein,
  title = {Density-functional theory and strong interactions: Orbital ordering in Mott-Hubbard insulators},
  author = {Liechtenstein, A. I. and Anisimov, V. I. and Zaanen, J.},
  journal = {Phys. Rev. B},
  volume = {52},
  issue = {8},
  pages = {R5467--R5470},
  numpages = {0},
  year = {1995},
  month = {Aug},
  publisher = {American Physical Society},
  doi = {10.1103/PhysRevB.52.R5467},
  url = {https://link.aps.org/doi/10.1103/PhysRevB.52.R5467}
}

@article{Cable1962,
  title = {Neutron-Diffraction Study of Antiferromagnetic Fe${\mathrm{Cl}}_{3}$},
  author = {Cable, J. W. and Wilkinson, M. K. and Wollan, E. O. and Koehler, W. C.},
  journal = {Phys. Rev.},
  volume = {127},
  issue = {3},
  pages = {714--717},
  numpages = {0},
  year = {1962},
  month = {Aug},
  publisher = {American Physical Society},
  doi = {10.1103/PhysRev.127.714},
  url = {https://link.aps.org/doi/10.1103/PhysRev.127.714}
}

@article{KOCHER1967,
title = {Mössbauer effect study of antiferromagnetic FeCl3},
journal = {Physics Letters A},
volume = {24},
number = {2},
pages = {93-94},
year = {1967},
issn = {0375-9601},
doi = {https://doi.org/10.1016/0375-9601(67)90498-7},
url = {https://www.sciencedirect.com/science/article/pii/0375960167904987},
author = {C.W. Kocher}
}

@article{LOBSTER,
	author = {Nelson, Ryky and Ertural, Christina and George, Janine and Deringer, Volker L. and Hautier, Geoffroy and Dronskowski, Richard},
	journal = {Journal of Computational Chemistry},
    doi = { https://doi.org/10.1002/jcc.26353},
	number = {21},
	pages = {1931-1940},
	title = {LOBSTER: Local orbital projections, atomic charges, and chemical-bonding analysis from projector-augmented-wave-based density-functional theory},
	volume = {41},
	year = {2020}}

@article{VASPKIT,
title = {VASPKIT: A user-friendly interface facilitating high-throughput computing and analysis using VASP code},
journal = {Computer Physics Communications},
volume = {267},
pages = {108033},
year = {2021},
doi = {https://doi.org/10.1016/j.cpc.2021.108033},
author = {Vei Wang and Nan Xu and Jin-Cheng Liu and Gang Tang and Wen-Tong Geng},
}

@article{Huang2017,
  title   = {Layer-dependent ferromagnetism in a van der Waals crystal down to the monolayer limit},
  author  = {Huang, B. and Clark, G. and Navarro-Moratalla, E. and Klein, D. R. and Cheng, R. and Seyler, K. L. and Zhong, D. and Schmidgall, E. and McGuire, M. A. and Cobden, D. H. and Yao, W. and Xiao, D. and Jarillo-Herrero, P. and Xu, X.},
  journal = {Nature},
  volume  = {546},
  pages   = {265--269},
  year    = {2017},
  doi     = {10.1038/nature22391}
}

@article{Song2018,
  title   = {Giant tunneling magnetoresistance in spin-filter van der Waals heterostructures},
  author  = {Song, T. and Cai, X. and Tu, M. W. Y. and Zhang, X. and Huang, B. and Wilson, N. P. and Seyler, K. L. and Zhu, L. and Taniguchi, T. and Watanabe, K. and McGuire, M. A. and Cobden, D. H. and Xiao, D. and Yao, W. and Xu, X.},
  journal = {Nature Nanotechnology},
  volume  = {13},
  pages   = {894--899},
  year    = {2018},
  doi     = {10.1038/s41565-018-0121-3}
}

@article{Jiang2018,
  title   = {Controlling magnetism in 2D CrI$_3$ by electrostatic doping},
  author  = {Jiang, S. and Li, L. and Wang, Z. and Mak, K. F. and Shan, J.},
  journal = {Nature Materials},
  volume  = {17},
  pages   = {406--410},
  year    = {2018},
  doi     = {10.1038/s41563-018-0040-6}
}

@article{Zhang2024_PRB,
  title     = {Two-dimensional magnetic materials with tunable topological properties},
  author    = {Zhang, Y. and Liu, H. and Wang, X. and Chen, Q.},
  journal   = {Physical Review B},
  volume    = {110},
  number    = {21},
  pages     = {214424},
  year      = {2024},
  publisher = {American Physical Society},
  doi       = {10.1103/PhysRevB.110.214424},
  url       = {https://doi.org/10.1103/PhysRevB.110.214424}
}

@article{LVTS12,
  title = {Wannier-based calculation of the orbital magnetization in crystals},
  author = {Lopez, M. G. and Vanderbilt, David and Thonhauser, T. and Souza, Ivo},
  journal = {Phys. Rev. B},
  volume = {85},
  issue = {1},
  pages = {014435},
  numpages = {12},
  year = {2012},
  month = {Jan},
  publisher = {American Physical Society},
  doi = {10.1103/PhysRevB.85.014435},
  url = {https://link.aps.org/doi/10.1103/PhysRevB.85.014435}
}

@article{Chang2013_QAH,
  author  = {Chang, Cui-Zu and Zhang, Jinsong and Feng, Xiao and Shen, Jie and Zhang, Zuocheng and Guo, Minghua and Li, Kang and Ou, Yunbo and Wei, Pang and Wang, Li-Li and Ji, Zhong-Qing and Feng, Yang and Ji, Shuaihua and Chen, Xi and Jia, Jinfeng and Dai, Xi and Fang, Zhong and Zhang, Shou-Cheng and He, Ke and Wang, Yayu and Lu, Li and Ma, Xu-Cun and Xue, Qi-Kun},
  title   = {Experimental observation of the quantum anomalous Hall effect in a magnetic topological insulator},
  journal = {Science},
  year    = {2013},
  volume  = {340},
  number  = {6129},
  pages   = {167--170},
  doi     = {10.1126/science.1234414},
  url     = {https://doi.org/10.1126/science.1234414}
}

@article{Chang2023_RMP,
  author  = {Chang, Cui-Zu and Liu, Chao-Xing and MacDonald, Allan H.},
  title   = {Colloquium: Quantum anomalous Hall effect},
  journal = {Reviews of Modern Physics},
  volume  = {95},
  pages   = {011002},
  year    = {2023},
  doi     = {10.1103/RevModPhys.95.011002},
  url     = {https://doi.org/10.1103/RevModPhys.95.011002}
}
%check file mybibitem.bib carefully!

\end{document}